\documentclass[twocolumn]{aastex62}

\newcommand{\afr}{$Af\rho$}
\newcommand{\phnm}{~per~0.1~\micron}

\graphicspath{{./}{figures/}}

\received{2018 November 30}
\revised{2018 December 28}
\accepted{2019 January 2}
\submitjournal{AJ}

\shorttitle{174P/Echeclus and its Blue Coma Observed Post-outburst}
\shortauthors{Seccull et al.}

\begin{document}

\title{174P/Echeclus and its Blue Coma Observed Post-outburst}

\correspondingauthor{Tom Seccull}
\email{tseccull01@qub.ac.uk}

\author[0000-0001-5605-1702]{Tom Seccull}
\affil{Astrophysics Research Centre, Queen's University Belfast, University Road, Belfast, BT7 1NN, UK}

\author[0000-0001-6680-6558]{Wesley C. Fraser}
\affiliation{Astrophysics Research Centre, Queen's University Belfast, University Road, Belfast, BT7 1NN, UK}

\author[0000-0003-0350-7061]{Thomas H. Puzia}
\affiliation{Institute of Astrophysics, Pontificia Universidad Cat\'olica de Chile, Av. Vincu\~na Mackenna 4860, 7820436, Santiago, Chile}

\author[0000-0003-0250-9911]{Alan Fitzsimmons}
\affiliation{Astrophysics Research Centre, Queen's University Belfast, University Road, Belfast, BT7 1NN, UK}

\author[0000-0002-6830-9093]{Guido Cupani}
\affiliation{INAF --- Osservatorio Astronomico di Trieste, via G. B. Tiepolo 11, I-34143, Trieste, Italy}



\begin{abstract}

It has been suggested that centaurs may lose their red surfaces and become bluer due to the onset of cometary activity, but the way in which cometary outbursts affect the surface composition and albedo of active centaurs is poorly understood. We obtained consistent visual-near-infrared (VNIR) reflectance spectra of the sporadically active centaur 174P/Echeclus during a period of inactivity in 2014 and six weeks after its outburst in 2016 to see if activity had observably changed the surface properties of the nucleus. We observed no change in the surface reflectance properties of Echeclus following the outburst compared to before, indicating that, in this case, any surface changes due to cometary activity were not sufficiently large to be observable from Earth. Our spectra and post-outburst imaging have revealed, however, that the remaining dust coma is not only blue compared to Echeclus, but also bluer than solar, with a spectral gradient of {$-7.7\pm0.6\%$\phnm}~measured through the $0.61-0.88~\micron$ wavelength range that appears to continue up to $\lambda\sim1.3~\micron$~ before becoming neutral. We conclude that the blue visual color of the dust is likely not a scattering effect, and instead may be indicative of the dust's carbon-rich composition. Deposition of such blue, carbon-rich, comatic dust onto a red active centaur may be a mechanism by which its surface color could be neutralized.

\end{abstract}



\section{Introduction} \label{sec:intro}

Centaurs are a population of minor planetary objects that currently reside amongst the outer planets of the solar system; their orbits are typically defined by a perihelion greater than Jupiter's semimajor axis ($q>5.2~au$) and a semimajor axis smaller than that of Neptune \citep[$a<30.1~au$;][]{2008ssbn.book...43G}. Most centaurs are thought to originate in the scattered disk of the trans-Neptunian belt, before gravitational interaction with Neptune forces their orbits to cross those of the giant planets \citep{1997Icar..127...13L,2004come.book..193D,2008ssbn.book..259G}. Such orbits are unstable with dynamical lifetimes of only the order of ${\sim}10^{6}-10^{7}$ yr \citep{1997Icar..127...13L,1999Icar..142..509D,2003AJ....126.3122T,2004MNRAS.354..798H,2007Icar..190..224D}. Gravitational planetary interactions can result in their ejection from the solar system \citep{2003AJ....126.3122T}, or the evolution of their orbits into those of Jupiter family comets \citep[JFCs;][]{2004come.book..193D}. As centaurs migrate toward the inner solar system, experiencing higher temperatures, some exhibit the onset of cometary activity. It has been suggested that cometary activity should cause the surfaces of red centaurs to change, such that they no longer look like red trans-Neptunian objects (TNOs), and instead look more similar to the neutrally colored JFCs \citep{1996AJ....112.2310L,2002AJ....123.1039J,2015AJ....150..201J,2009Icar..201..674L}.

Around 13\% of known centaurs have been observed to show cometary activity \citep{2009AJ....137.4296J}. From in-situ observations of the active JFC 67P/Churyumov-Gerasimenko (hereafter 67P) made by the European Space Agency's \textit{Rosetta} spacecraft, it is known that cometary activity changes the surface properties of a cometary nucleus. \citet{2016Icar..274..334F} reported that at 67P \textit{Rosetta's} VIRTIS instrument observed the single scattering albedo of active areas increasing, and their spectral slopes decreasing, suggesting that cometary outgassing was lifting redder dust from the surface to reveal more reflective and bluer subsurface water ice. The increase in surface water ice abundance was also observed in reflectance spectra of the active regions by the increasing depth of an absorption feature observed at $\lambda\sim3.2~\micron$, and the distortion of its center toward shorter wavelengths \citep{2016Icar..274..334F}. While these changes are clear in the measurements of 67P, they have not yet been  observed on the surface of an active centaur. 

Plausible evidence for centaur surface changes due to activity come from hemispherically averaged color measurements. The colors (or spectral slopes) of centaurs are bimodally distributed into a red group and a less-red group \citep{2003A&26A...410L..29P,2005AJ....130.1291B,2008ssbn.book..105T,2010A&26A...510A..53P} but active centaurs have only been observed within the less-red group \citep{2009AJ....137.4296J}. Red surfaces are also reported to be less common on centaurs with perihelia below ${\sim}10$~au; this heliocentric distance roughly coincides with that where activity is observed to begin \citep{2015AJ....150..201J}. It has been argued that activity can destroy the original red irradiated crust if a centaur and resurface it with less-red unirradiated subsurface material that falls back from the coma under gravity \citep{2002AJ....123.1039J,2004A&26A...417.1145D,2005Icar..174...90D}. Despite this apparent trend for activity causing  bluing of a centaur's spectrum, no changes in their surface spectral properties have been directly detected following observed activity.

174P/Echeclus \citep[also known as (60558) Echeclus, formerly known as 2000~EC$_{98}$;][ and hereafter simply referred to as Echeclus]{MPEC2000-E64} is a centaur that has been extensively studied while both inactive and active. Its orbit ($a=10.68$~au, $e=0.46$, $i=4.34\degr$, $q=5.82$~au, $t_{q}=$2015 April 22)\footnote{Orbital elements from the IAU Minor Planet Center} has been described as Jupiter coupled by \citet{2008ssbn.book...43G}, whose numerical integrations found Echeclus to be rapidly perturbed by Jupiter. Echeclus is in the less-red group of centaurs, and slope measurements obtained for the featureless visual spectrum of its bare nucleus are consistent at {${\sim}10-13\%$\phnm}  \citep{2002A&26A...395..297B,2003Icar..166..195B,2008A&26A...487..741A,2008ssbn.book..161S,2015A&26A...577A..35P}. A near-infrared (NIR) spectrum of Echeclus reported by \citet{2009Icar..201..272G} has a low spectral slope of {$2.53\pm0.33\%$\phnm}, and no water ice absorption features.

Since its discovery Echeclus has been observed outbursting four times. The first, discovered by \citet{IAUC8656}, was a large outburst that occurred in 2005 December. Two minor outbursts occurred in 2011 May \citep{IAUC9213} and 2016 August \citep{CBET4313}. Most recently another large outburst was discovered in 2017 December by amateur astronomers\footnote{\url{https://www.britastro.org/node/11931}}.
Multiple works have been published on Echeclus' outburst of 2005-2006, which was one of the largest centaur outbursts ever observed. Persisting for several months, it rose in visual magnitude from $\sim$21 to $\sim$14 \citep{2008A&26A...480..543R}. Unusually, the source of activity appeared to be distinct from Echeclus itself, and was possibly a fragment of the nucleus broken off by the outburst \citep{CBET563,BAAS38551,2008PASP..120..393B,2008A&26A...480..543R,2009P&26SS...57.1218F}. Both \citet{2008PASP..120..393B} and \citet{2008A&26A...480..543R} reported high dust production rates, low gas--dust ratios, and dust colors redder than solar but more neutral than Echeclus itself. These dust color properties were also apparent in observations of Echeclus' 2011 outburst \citep{2016MNRAS.462S.432R} and have been attributed to dust grain size and scattering effects \citep{2008PASP..120..393B,2008A&26A...480..543R,2016MNRAS.462S.432R}. While Echeclus' outbursts \citep[and CO outgassing;][]{2017AJ....153..230W} have received much attention, any effects of an outburst on this centaur's surface composition have not been reported.

\begin{table*}
\begin{center}
\caption{Observation Geometry for Echeclus}
\label{tab:obsgeom}
\begin{tabular}{ p{2.5cm}p{1.5cm}p{1.6cm}p{1.3cm}p{1.1cm}p{0.9cm}p{1.6cm}p{1.8cm}p{2.0cm}}
\hline\hline
Observation Date & R.A. (hr) & Decl. (deg) & $r_{H}$\textsuperscript{a} (au) & $\Delta$\textsuperscript{b} (au) & $\alpha$\textsuperscript{c}($\degr$) & PsAng\textsuperscript{d} ($\degr$) & PsAMV\textsuperscript{e} ($\degr$) & Days after $t_{q}$\textsuperscript{f} \\[2pt]
\hline
Pre-outburst & & & & & & & &\\
\hline
2014 Aug 3 & 21:06:49.21 & -13:07:20.4 & 5.952 & 4.941 & 0.9623 & 217.377 & 256.494 & -262.4\\[1pt]
\hline
Post-outburst & & & & & & & &\\
\hline
2016 Oct 7 & 01:16:09.46 & +05:39:09.2 & 6.343 & 5.348 & 0.9490 & 269.594 & 251.716 & 533.5\\[1pt]
\hline
2016 Oct 8 & 01:15:48.15 & +05:36:36.6 & 6.344 & 5.348 & 0.7962 & 274.173 & 251.704 & 534.5\\[1pt]

\hline
\end{tabular}\\[6pt]
\small{\textbf{Note.} Values reported in this table correspond to observation geometry at the median time of each set of spectroscopic observations reported in Table \ref{tab:obsdets}. (a) heliocentric distance, (b) geocentric distance, (c) solar phase angle, and (d) and (e) position angles of the extended Sun-to-Echeclus radius vector and the negative of Echeclus' heliocentric velocity vector as seen in the observer's plane of sky (both are plotted in Fig. \ref{fig:f1}); these angles are measured counterclockwise (eastward) relative to the northward direction. (f) Number of days after Echeclus passed perihelion that the observation took place.}
\end{center}
\end{table*}


\begin{table*}
\begin{center}
\caption{Spectroscopic Observation Log}
\label{tab:obsdets}
\begin{tabular}{ p{2.4cm}p{4.8cm}p{0.8cm}p{0.8cm}p{0.9cm}p{1.6cm}p{2cm}p{1.8cm}}
\hline\hline
Target & Observation Date $\vert$ UT Time & \multicolumn{3}{c}{Exposure Time (s)} & Exposures & Air Mass & Seeing (\arcsec) \\[2pt]
\hline
Pre-outburst & & UVB & VIS & NIR & & &\\
\hline
HD 198289 & 2014 Aug 3 $\vert$ 04:21:28--04:27:35 & 3.0 & 3.0 & 12.0 & 3 & 1.020--1.023 & 0.66--0.89 \\[1pt]
HD 198289 & 2014 Aug 3 $\vert$ 04:31:56--04:38:05 & 10.0 & 10.0 & 40.0 & 3 & 1.018--1.019 & 0.68--0.76 \\[1pt]
Echeclus & 2014 Aug 3 $\vert$ 05:16:05--07:24:17 & 500.0 & 466.0 & 532.0 & 12 & 1.022--1.259 & 0.81--1.16 \\[1pt]
HIP 107708 & 2014 Aug 3 $\vert$ 07:50:27--07:56:29 & 6.0 & 6.0 & 24.0 & 3 & 1.141--1.161 & 1.02--1.24 \\[1pt]
HIP 105408 & 2014 Aug 3 $\vert$ 08:04:25--08:10:55 & 10.0 & 10.0 & 40.0 & 3 & 1.304--1.329 & 1.21--1.41 \\[1pt]

\hline
Post-outburst & & & & & &\\
\hline
HIP 107708 & 2016 Oct 7 $\vert$ 04:29:03--04:35:02 & 6.0 & 6.0 & 24.0 & 3 & 1.336--1.371 & 0.66--0.75 \\[1pt]
Echeclus & 2016 Oct 7 $\vert$ 04:53:03--05:51:00 & 500.0 & 466.0 & 532.0 & 6 & 1.159--1.211 & 0.63--0.70 \\[1pt]
HD 16017\textsuperscript{a} & 2016-10-07 $\vert$ 06:20:03--06:23:02 & 6.0 & 6.0 & 24.0 & 2 & 1.154--1.155 & 0.45--0.55 \\[1pt]

\hline
HIP 107708 & 2016 Oct 8 $\vert$ 04:04:17--04:35:02 & 6.0 & 6.0 & 24.0 & 3 & 1.249--1.278 & 0.66--0.83\\[1pt]
Echeclus & 2016 Oct 8 $\vert$ 04:35:56--05:33:47 & 500.0 & 466.0 & 532.0 & 6 & 1.158--1.191 & 0.73--0.88 \\[1pt]
HD 16017 & 2016 Oct 8 $\vert$ 05:53:55--05:59:56 & 6.0 & 6.0 & 24.0 & 3 & 1.154--1.156 & 0.72--0.81 \\[1pt]

\hline
\end{tabular}\\[6pt]
\small{\textbf{Note.} \textsuperscript{a} This triplet of exposures could not be completed due to time constraints.}
\end{center}
\end{table*}

In 2014 August we obtained a high-quality visual-NIR (VNIR) spectrum of Echeclus' bare nucleus as part of a program observing centaurs and TNOs with the X-Shooter spectrograph \citep{2011A&A...536A.105V} mounted on the European Southern Observatory (ESO) Very Large Telescope (VLT). Echeclus unexpectedly outburst on 2016 August 27 at a heliocentric distance of 6.27~au; it increased in brightness by 2.6 mag in r' in less than a day \citep{CBET4313}. Additional spectra of similarly high quality were gathered in 2016 October, six weeks after the outburst, to enable a direct measure of any new absorption features, changes in Echeclus' visual and NIR spectral slopes, and the color of any residual dust coma. We discuss both sets of observations, and provide interpretation below.

\section{Observations} \label{sec:obs}
Echeclus was observed during two epochs (before and after the 2016 outburst) using the X-Shooter spectrograph at the ESO VLT. Pre-outburst observations were performed in visitor mode on 2014 August 3, and post-outburst observations were performed in service mode over two nights on 2016 October 7-8. Our observations were designed to be the same in both epochs; this is the case unless otherwise specified.

X-Shooter is a medium resolution echelle spectrograph with three arms that can be exposed simultaneously, covering the 
near-UV/blue (UVB; 0.30---0.56~\micron), visual (VIS; 0.55---1.02~\micron), and NIR (1.02---2.48~\micron) spectral ranges  \citep{2011A&A...536A.105V}. While our pre-outburst observations used X-Shooter's full wavelength coverage, we used its \textit{K}-band blocking filter for our post-outburst observations. This filter blocks incoming flux at wavelengths longer than 2.1~\micron~while boosting the signal-to-noise ratio (S/N) in the rest of the NIR spectrum. The UVB, VIS, and NIR detectors have respective pixel scales of 0.16\arcsec, 0.16\arcsec~and 0.21\arcsec. We set the UVB, VIS, and NIR slits to widths of 1.0\arcsec, 0.9\arcsec~and 0.9\arcsec, each providing a respective resolving power of 
$\sim$5400, $\sim$8900, and $\sim$5600. The slits for each arm have a common length of 11\arcsec. No readout binning was performed for any of the observations.

The spectra were observed in a three-point dither pattern to mitigate 
bad pixel artifacts and cosmic-ray contamination. The 
slit was realigned to the parallactic angle at the beginning of each triplet to 
reduce the effects of atmospheric differential refraction; this was especially 
important because X-Shooter's atmospheric dispersion corrector (ADC) was disabled during both observing runs.
During the pre-outburst run three solar calibrator stars were observed with the same instrument setup, adjacent in time to Echeclus, and at 
similar air mass: these included HD~198289, Hip~107708, 
and Hip~105408. Similarly, two calibrator stars were observed on each night during the post-outburst run, Hip~107708 and HD~16017. We ensured that at least one star, Hip~107708, was common to both observing runs so a direct comparison could be made between the pre- and post-outburst reflectance spectra. During the pre-outburst run, flux calibrators EG~274 and Feige~110 were observed as part of the 
standard X-Shooter calibration program. Similarly, flux standards LTT~3218 and LTT~7987 were observed during the post-outburst run \citep{2010HiA....15..535V}. The observation geometry for Echeclus and a spectroscopic observation log are respectively reported in Tables~\ref{tab:obsgeom}~\&~\ref{tab:obsdets}.

\begin{figure*}
\includegraphics[scale=0.46]{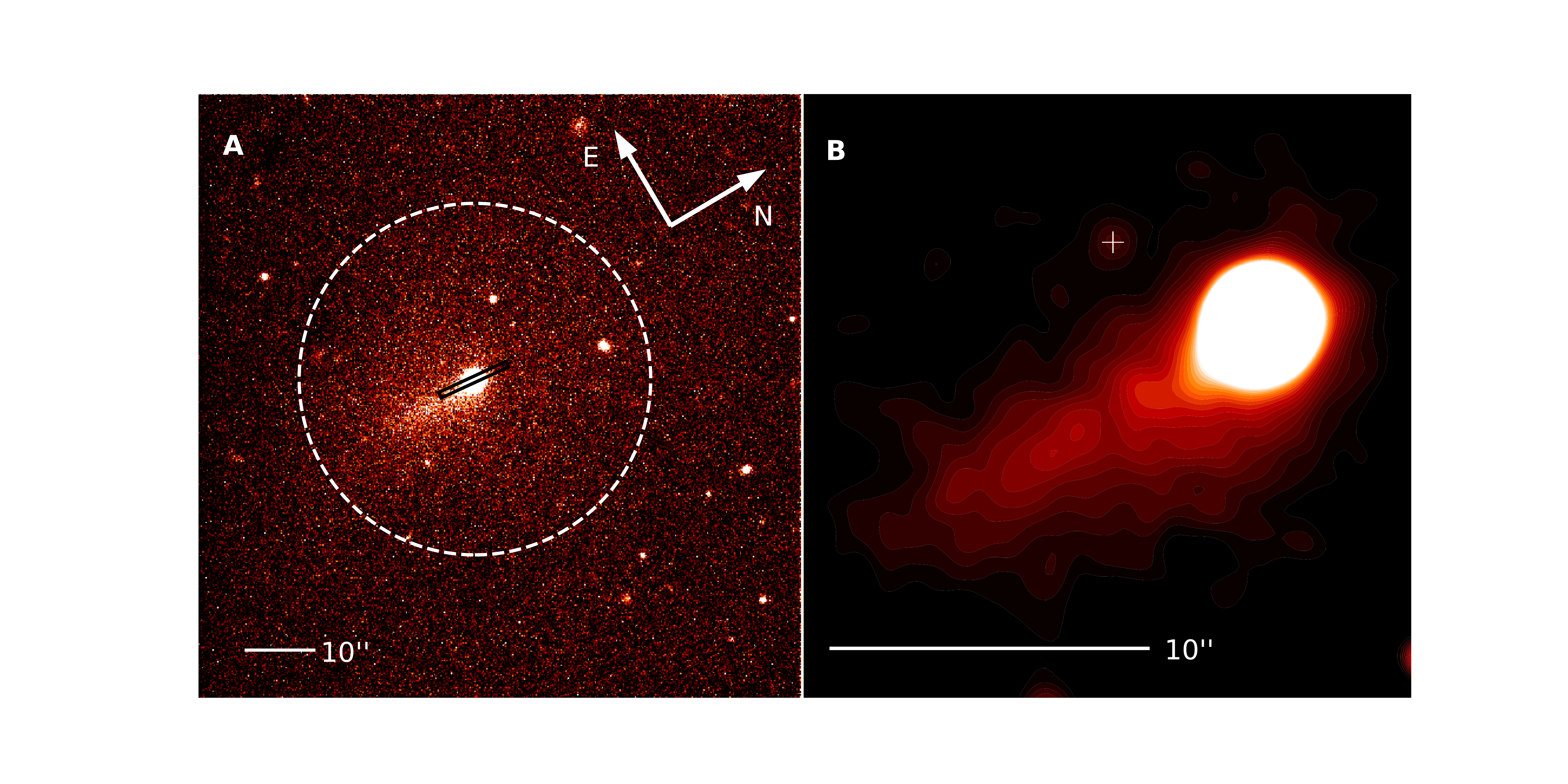}
\caption{Panel (A) shows a debiased, flat-fielded, 45 second \textit{r'} exposure of Echeclus and its coma, observed with the AGC of X-Shooter on 2016 October 7. The dashed ring has a radius of 26\arcsec~($\sim1\times10^{5}$~km at Echeclus) and is centered on the nucleus, marking the point at which the radially averaged surface brightness of the coma blends into that of the background sky. The small black rectangle marks the size and average orientation of the slit of X-Shooter while obtaining spectra of Echeclus on 2016 October 7. Arrows labeled N and E respectively mark the north and east directions on the sky; arrows labeled PsAng and PsAMV respectively point in directions opposite to the position of the Sun on the sky, and opposite to the heliocentric orbital motion of Echeclus (angle values are reported in Table \ref{tab:obsgeom}). Panel (B) shows a zoomed contour plot of the same image after it was smoothed with a gaussian filter. It has linear scaling, and shows the observed morphology of Echeclus' coma, which is similar to that observed in our \textit{g'} image on the same night. The lowest contour is set at one standard deviation of the background noise above the median background level. The + symbol marks a background source that is unrelated to Echeclus.  \label{fig:f1}}
\end{figure*}

X-Shooter also has a limited imaging mode \citep{2014Msngr.156...21M}, which we used to study Echeclus' coma. This imaging mode uses X-Shooter's acquisition and guiding camera (AGC) which has a $512\times512$ pixel E2V broadband coated CCD, a $1.5\arcmin\times1.5\arcmin$ field of view, a pixel scale of $\sim$0.17\arcsec~and a number of standard photometric filters. On the first night of the post-outburst observing run three images were obtained in both Sloan Digital Sky Survey (SDSS) \textit{g'} and \textit{r'} filters, one each of exposure lengths 5s, 15s, and 45s (see Fig. \ref{fig:f1}).     

\section{Data Reduction} \label{sec:red}
\subsection{Spectra: Echeclus} \label{sec:red1}
Standard reduction steps (including rectification and merging of the orders and flux calibration) were performed for all the observed Echeclus and calibrator spectra with the ESO X-Shooter data reduction pipeline \citep[v. 2.7.1;][]{2010SPIE.7737E..28M} in the ESO Reflex data processing environment \citep[v. 2.8.4;][]{2013A&A...559A..96F}. Sky subtraction, cosmic-ray removal and extraction of Echeclus' spectrum were performed in a similar manner to that described in more detail by \citet{2018ApJ...855L..26S}. Briefly,  Moffat functions \citep{1969A&A.....3..455M} were fitted to the spatial profile of the 2D rectified and merged spectrum at many locations along the dispersion axis in order to track the wavelength dependent center and width of the spectrum's spatial profile. Sky region boundaries were defined at $\pm3$ Full-Widths at Half Maximum (FWHM) from each Moffat profile center with sky outside these boundaries. Cosmic rays in the sky and target regions were separately sigma clipped at 5$\sigma$. In each unbinned wavelength element the sky and dust flux contribution was subtracted with a linear fit. Another round of Moffat fitting was conducted for the sky-subtracted spectrum and extraction limits were defined at $\pm2$ FWHM from each of the Moffat profile centers. Within these limits the flux was summed for each wavelength element to form the 1D spectrum. Following extraction, the individual spectra were median stacked, solar 
calibrated, and binned. In each bin the data points were sigma clipped at $3\sigma$ to minimize skew from outlying points; the remaining points were bootstrapped $10^{3}$ times to produce a distribution of medians and standard deviations for the bin. The mean of the distribution of medians was taken as the bin value and the mean of the distribution of standard deviations was used to calculate the standard error of the bin, which was taken as the bin's uncertainty. Dithers with low S/N or large residuals following sky subtraction were omitted from the final stack.

While the above extraction method is good for extracting spectra of faint sources \citep[e.g.][]{2018NatAs...2..133F,2018ApJ...855L..26S}, it cuts off the overlapping ends of each spectral arm where X-Shooter's dichroics reduce the S/N of the spectrum to a point where it is too low to properly extract; this makes alignment of spectra in neighboring arms a non-trivial task in some cases. The UVB and VIS spectra were simply aligned via a linear fit through the spectral ranges adjacent to the join between them. In all of the VIS arm reflectance spectra there is a strong telluric residual feature at 0.9--0.98~\micron~\citep[see][]{2015A&26A...576A..77S}. This produced a large enough gap in usable continuum between the VIS and NIR spectra that we chose not to attempt to align the VIS and NIR arms. Hence, in the following analysis, we consider the combined UVB-VIS spectrum, and the NIR spectrum separately. 

All of the spectra have been cut below 0.4~\micron~due to strong residuals caused by differences in metallicity between the Sun and the calibrator stars used \citep[see][]{1980A&A....91..221H}.  The spectra have also been cut above 2.1~\micron~where the sky subtraction was very poor. Direct comparison of the pre- and post-outburst spectra is also not possible at wavelengths above 2.1~\micron~due to the K-blocking filter used in the post-outburst observations.

\subsection{Spectra: Dust}

As discussed in Section \ref{sec:obs}, Echeclus was observed in a three-point dither pattern. In the first dither position Echeclus was centered in the slit. In the second and third dithers Echeclus was positioned respectively at +2.5\arcsec and -2.5\arcsec~along the slit relative to the center. Due to the asymmetry of the coma  and the convenient orientation of the slit on the sky during the post-outburst observations, one end of the slit sampled the dust coma while the other was dominated by sky (see Fig. \ref{fig:f1}). This made the VNIR extraction of a dust spectrum possible. To extract the dust spectrum we used only the spectra acquired at the second and third  dither positions, to extract the dust and the sky spectra, respectively.

The post-outburst 2D rectified and merged spectra produced by the ESO pipeline, and the sky boundaries drawn by the extraction process described in Section \ref{sec:red1} were used to sky-subtract and extract the spectrum of the coma. The sky region for each dust spectrum was defined in the third dither in a triplet, on the dust-free side of the spectral image at \textgreater3 FWHM from the center of Echeclus' spatial profile. Likewise, the dust region in each triplet was defined in the second dither, on the dusty side of the spectral image at \textgreater3 FWHM from the center of Echeclus' spatial profile. A sky spectrum was produced for each triplet by taking a median of the pixels in each wavelength element in the 2D sky region. This sky spectrum was then subtracted from the second dither 2D spectrum, and pixels in each wavelength element in the dust region were summed to produce a sky-subtracted 1D dust spectrum.  As a result of summing only the flux at $>$3 FWHM from Echeclus' nucleus the contamination of the dust spectrum by that of Echeclus itself is expected to be minimal.

Like the spectra of the nucleus, the four resulting dust spectra were stacked, solar calibrated, and binned. The uncertainties in each spectral bin were determined in the same way as for those in the spectrum of Echeclus (see Section \ref{sec:red1}). The dust reflectance spectrum appears to have a nearly linear behavior through the VNIR range (from $\sim$0.4-1.3 \micron; see Fig. \ref{fig:f3}) and so we were able to align the UVB, VIS, and NIR spectra by aligning the UVB and NIR spectra to a linear fit of the VIS spectrum. The dust spectrum was cut above 1.76~\micron~where S/N is extremely low.

\subsection{Images} \label{sec:imred}
There is no dedicated data reduction pipeline for images observed by the AGC of X-Shooter, so we debiased and flat-fielded the images using custom scripts written in Python v. 2.7.13\footnote{\url{https://www.python.org/}}. Bias frames were median stacked to make a master bias frame, which was subtracted from the sky flats and the image. The bias-subtracted sky flats were averaged, normalized, and divided from the image to flat-field it. It should be noted that the quality of the flat field was moderately poor towards the edges of the image. The calibrated image was sky-subtracted with TRIPPy \citep{2016AJ....151..158F}.

From the post-outburst images we aimed to study Echeclus' remaining dust coma, and to maximize the S/N we chose to only use those with the longest exposure time of 45s in the \textit{g'} and \textit{r'} filters. Due to the very small field of view of X-Shooter's AGC ($1.5\arcmin\times1.5\arcmin$), we were limited in the number of sources we could use to calculate the zero-point of the images. Cross-referencing them with the catalogs of the SDSS \citep[][]{2018ApJS..235...42A} and the Panoramic Survey Telescope and Rapid Response System \citep[Pan-STARRS; ][in preparation]{arXiv1612.05243} revealed that they contained only one usable stellar source, as all others beside Echeclus were galaxies. In \textit{g'} the uncertainty in the star's magnitude was significant ($\sim0.13$ mag) and an accurate zero-point for this image could not be determined; as a result the colors determined for Echeclus and its coma also have significant uncertainty. To determine a lower limit on the mass of the coma we were limited to using only the \textit{r'} image. The uncertainty in all subsequent photometric measurements is dominated by the uncertainty in the measured zero-points.

\section{Results and Analysis}
\begin{figure*}
\includegraphics[scale=1.]{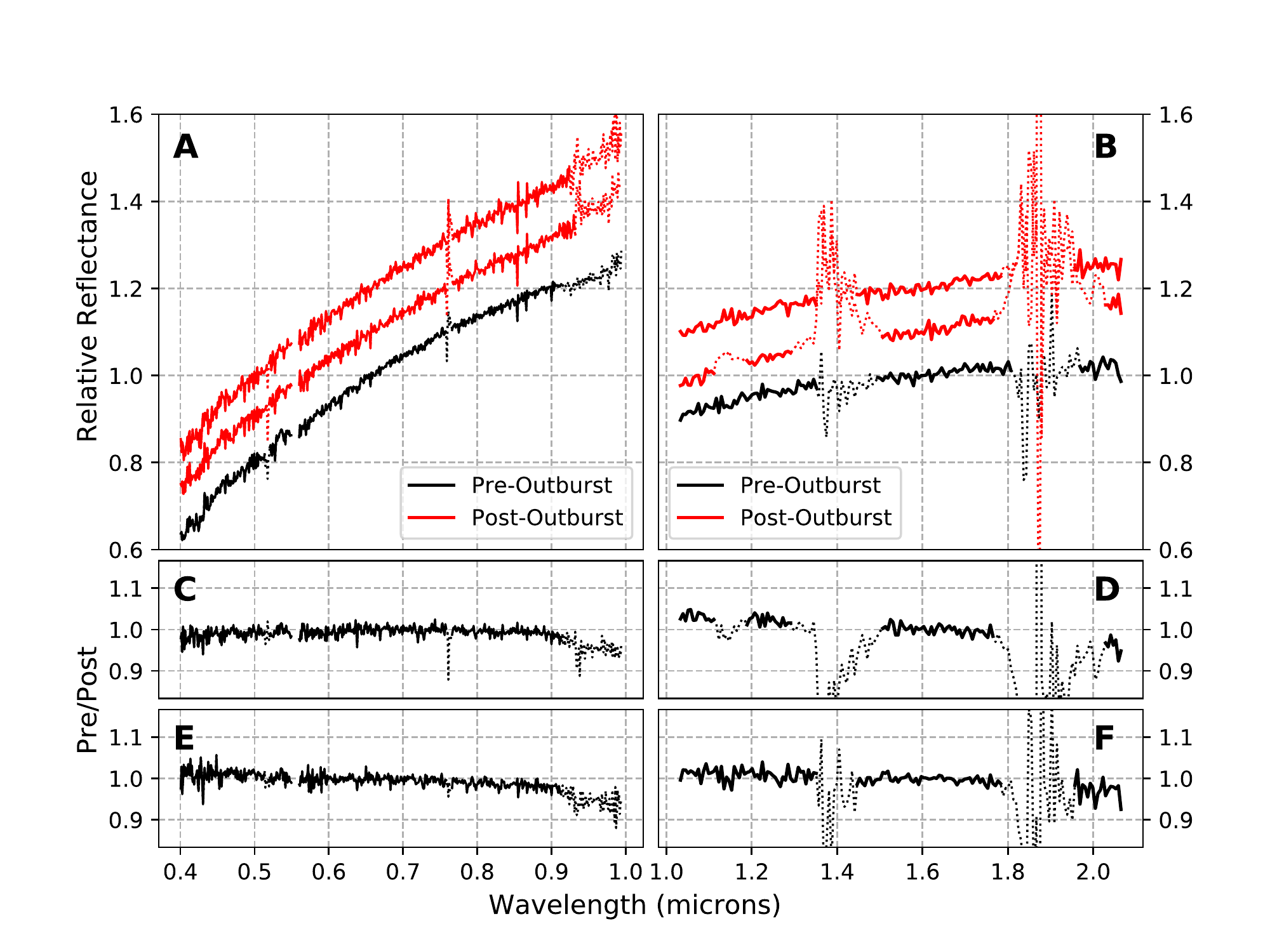}
\caption{Reflectance spectra  and ratioed spectra of Echeclus' surface from both observing epochs. All of panels on the left, (A), (C), and (E), show the visual range and are normalized at 0.658~\micron, while those on the right, (B), (D), and (F), display the NIR and are normalized at 1.6~\micron. In all of the panels the y-axis scaling is the same, and spectral regions contaminated by telluric or solar metallicity residuals are plotted with dotted lines. In panels (A) and (B) the reflectance spectra of Echeclus' surface are shown, with each spectrum ordered from bottom to top in order of when they were observed, and offset for clarity by +0.1 with respect to the previous spectrum. During both observing epochs the visual spectrum of Echeclus is featureless and no statistically significant change in the spectral gradient is observed; the same is observed in the NIR. The differing strength and width of the telluric residual bands are related to the difference in air mass at which Echeclus and the calibrator star were observed on a given night (see Table \ref{tab:obsdets}). All of the spectra displayed have been calibrated using the spectrum of Hip 107708. The ratioed spectra in panels (C) and (D) were created by dividing the reflectance spectra from 2014 August 3 by those from 2016 October 7.  The same applies to panels (E) and (F) where we used spectra from 2014 August 3 and 2016 October 8. See Section \ref{sec:echspec} for discussion of the ratioed spectra. The data used to create this figure are available with the AJ article, or may be requested from T.S. \label{fig:f2}}
\end{figure*}

\begin{figure*}
\includegraphics[scale=1.0]{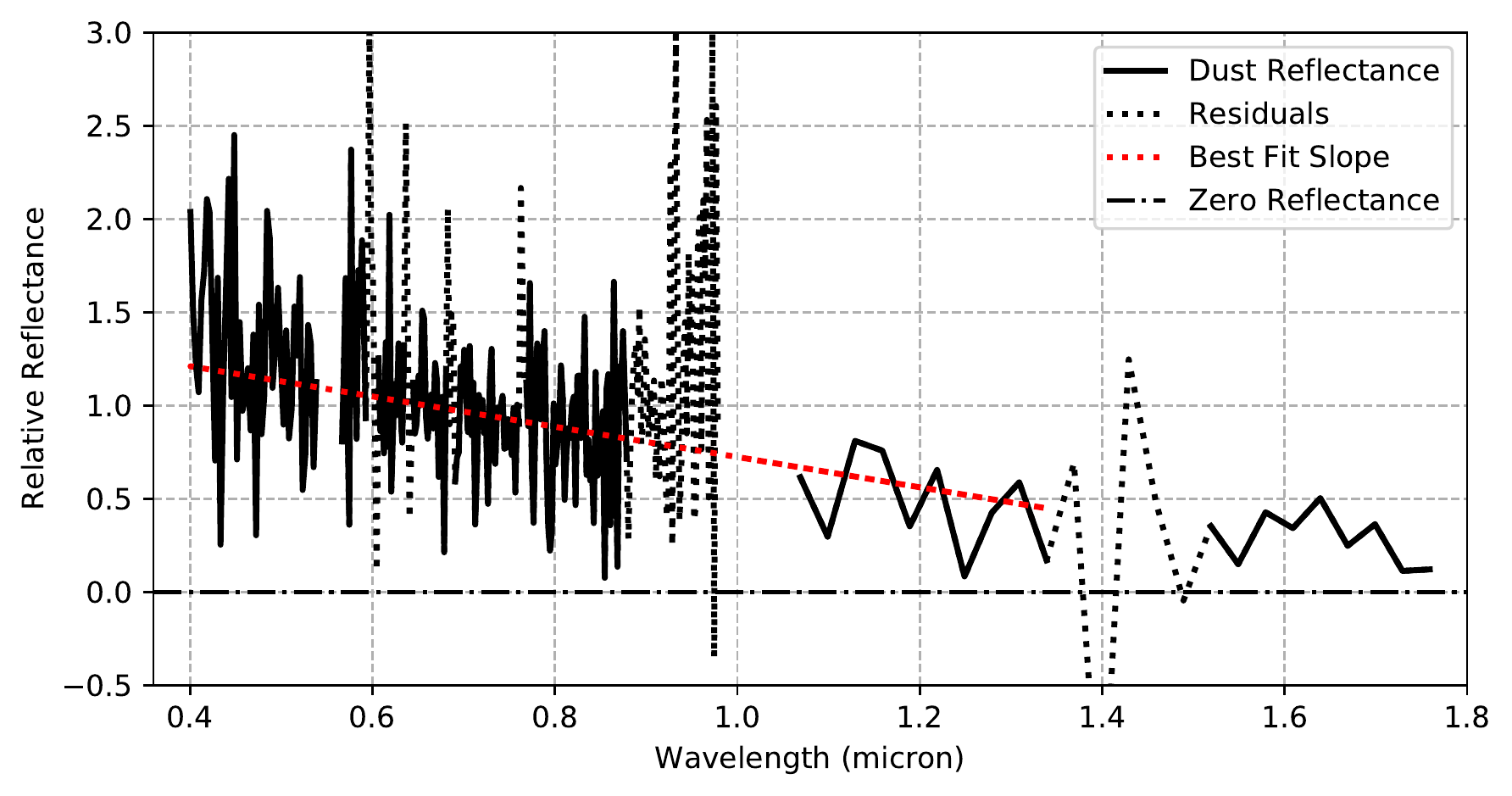}
\caption{The blue, featureless reflectance spectrum of Echeclus' dust coma observed post-outburst. The spectrum is normalized at 0.658~\micron. The dotted lines indicate regions of telluric residuals and systematic residuals produced by very low S/N at the ends of X-Shooter's echelle orders. The red dotted line plots the measured best-fit slope of {$-7.7\pm0.6\%$\phnm} across the 0.40-1.35~\micron~range (see Section \ref{sec:resdust}). The black dotted-dashed line at zero reflectance indicates the lower limit of the plot where reflectances are still physical. This spectrum was also calibrated using that of Hip 107708. The data used to create this figure are available with the AJ article, or may be requested from T.S. \label{fig:f3}}
\end{figure*}
For Echeclus we obtained 3 reflectance spectra in both the visual (combined UVB+VIS arms) and NIR ranges (see Fig. \ref{fig:f2}). In each range we obtained one pre-outburst baseline spectrum and two post-outburst spectra. We also obtained one full VNIR spectrum of Echeclus' dust coma from the combined post-outburst observations (see Fig. \ref{fig:f3}). Prior to the solar calibration of our data we searched the spectra for cometary emission lines that might have indicated ongoing activity, but found none above the level of the noise.

In each reflectance spectrum we measured the spectral gradient, $S'$. For each point in a binned spectrum we took its constituent set of unbinned values; these values were bootstrapped allowing for repeats, and were then rebinned. This was done $10^{3}$ times for every point in the binned spectrum to produce $10^{3}$ sample binned spectra, on each of which a linear regression was performed to create a distribution of $10^{3}$ spectral slopes. $S'$ was defined as the mean of this slope distribution; standard error of the mean at 99.7\% confidence was also recorded (see Table \ref{tab:slopes}). The $S'$ standard errors are very small ({$0.01\%$\phnm}) and do not account for systematic errors of order {$0.5\%$\phnm} introduced by imperfect calibration of the data; these systematics can be seen in the curvature and slopes of the ratioed spectra displayed in Fig. \ref{fig:f2}. Hence we report the measured uncertainty of {$0.01\%$\phnm} as the limit of precision obtainable from our continuum measurements of these spectra. The  uncertainties quoted in Table 2. do not include the systematic errors.
\newpage

\subsection{Spectra: Echeclus} \label{sec:echspec}
We observed no appreciable change in the reflectance properties of Echeclus' nucleus post-outburst compared to our pre-outburst baseline spectrum; no observable ice or silicate absorption features have appeared, nor has the shape of the spectrum itself altered.

Ratioed spectra were produced to probe for changes in Echeclus' reflectance properties (see Fig. \ref{fig:f2}). They show very slight curvature in the visual range. At longer wavelengths they have a small negative gradient. These residual features are not significant, producing maximum residual spectral gradients of order {$0.5\%$\phnm}. These residual features are likely not intrinsic to Echeclus, but instead are the result of strong telluric residual contamination and systematic errors introduced by imperfect calibration.

\begin{table}
\caption{Reflectance Spectrum Gradients}
\label{tab:slopes}
\begin{tabular}{p{2.2cm}p{2.3cm}p{2.4cm}}
\hline\hline
Obs Date & $S'$ ({\%\phnm}) & Standard Error\textsuperscript{a} ({\%\phnm}) \\[1pt]
\hline
Visual\\[2pt]
\hline
2014 Aug 3 & 11.64 & $\pm0.01$ \\[1pt]
2016 Oct 7 & 11.39 & $\pm0.01$ \\[1pt]
2016 Oct 8 & 12.12 & $\pm0.01$ \\[1pt]
\hline
NIR\\[1pt]
\hline
2014 Aug 3 &1.26 & $\pm0.01$ \\[1pt]
2016 Oct 7 & 1.69 & $\pm0.01$ \\[1pt]
2016 Oct 8 & 1.34 & $\pm0.01$ \\[1pt]
\hline
\end{tabular}\\[6pt]
\small{\textbf{Notes.} Calculated with VIS and NIR reflectance respectively normalized at 0.658 \micron~and 1.6 \micron.\newline \textsuperscript{a} to account for systematic errors increase these values by {$0.5\%$\phnm}}
\end{table}

In the visual spectra the pre-outburst slope was measured at 0.575-0.800~\micron~while ignoring the telluric residuals at 0.758-0.767~\micron. Visual $S'$ values and their uncertainties are presented in Table \ref{tab:slopes}. The visual spectral gradients reported here are all consistent with literature values previously published for Echeclus when systematic errors are accounted for \citep{2008A&26A...487..741A,2008ssbn.book..161S,2015A&26A...577A..35P}. 

To enable direct comparison between our NIR $S'$ measurements, we measured the slope in the wavelength range of 1.25-1.7~\micron, but only in regions with minimal telluric residual contamination due to atmospheric H$_{2}$O \citep[see ][]{2015A&26A...576A..77S}. Hence, we only included data in the ranges of 1.25-1.3~\micron~and 1.5-1.7~\micron~in our measurements. Outside of these regions even subtle telluric residuals were able to affect $S'$, despite their apparent absence from the spectrum. At $\lambda < 1.2$~\micron~the slope of Echeclus' NIR spectrum begins to increase toward the visual range and drops away from linearity; this is why we did not include these wavelengths in our measurements of $S'$. NIR $S'$ values and their uncertainties are also presented in Table \ref{tab:slopes}. The NIR spectral gradients are broadly neutral like that reported by \citet{2009Icar..201..272G}, but our measured values are not formally consistent with theirs. Due to the significant difference between the wavelength range measured in that study and this one, however, they are not directly comparable. 

Taking into account both the formal and systematic uncertainties in our measurements of $S'$, we cannot report any changes in observed visual or NIR spectral gradients between the pre- and post-outburst epochs that are attributable to changes in the reflectance properties of Echeclus.
\newpage
\subsection{Spectra: Dust} \label{sec:resdust}
The reflectance spectrum of Echeclus' dust coma appears entirely featureless, but surprisingly 
blue in the visual range. Using the same method outlined above we measured the spectrum's gradient, $S'$, in the highest S/N region of the VIS dust spectrum (0.61-0.88~\micron), while ignoring regions containing telluric and calibration residuals. We measured the gradient of the VIS dust spectrum and standard error at 99.7\% confidence to be {$-7.7\pm0.4\%$\phnm}. Adding {$0.5\%$\phnm} to the slope uncertainty in quadrature to account for systematic error results in a final slope measurement of {$-7.7\pm0.6\%$\phnm}. The spectrum is broadly linear and appears to have a constant gradient from 0.4 to 1.3~\micron, while leveling out at longer wavelengths.

\subsection{Images: Colors}
From our \textit{g'} and \textit{r'} images we attempted to study the dust coma of Echeclus. Following the calibrations described in Section \ref{sec:imred}, the locations of sources in the image were found using SExtractor \citep{1996A&AS..117..393B} and using the single stellar source available, a stellar point-spread function (PSF) was fitted by TRIPPy \citep{2016AJ....151..158F}; TRIPPy was also used for all subsequent photometric measurements. The pixel values of three bright sources close to Echeclus in the image were replaced with values randomly sampled from a gaussian distribution calculated from the median and standard deviation of the background local to each star; this removed these sources from the image while preserving the background coma.

We measured the $g'-r'$ colors of Echeclus and its coma. First, to measure the color of Echeclus we performed aperture photometry with a circular aperture of radius, $\rho=0.33\arcsec$, equal to $\frac{1}{2}\times$FWHM of the seeing disk. This aperture was chosen to probe the color of the nucleus and minimize coma contamination. Second, to measure the color of the coma, photometry was performed for a ring centered on the photocenter of Echeclus with an inner bound at $\rho=1.98\arcsec$~($3\times$FWHM of Echeclus' seeing disk) and an outer bound at $\rho=26\arcsec$; this included most of Echeclus' observable coma, but minimized contamination by flux from the nucleus. For Echeclus and its coma we respectively measured $g'-r'$ colors of $0.65\pm0.14$ and $0.29\pm0.14$, where the uncertainties are dominated by the uncertainty introduced by the zero-point calibration of the \textit{g'} image. Despite the large uncertainty in these color measurements, the colors of Echeclus and its dust coma are not consistent. Importantly, the coma's color is neutral-blue (the Sun has $g'-r'=0.44\pm0.02$)\footnote{\url{https://www.sdss.org/dr12/algorithms/ugrizvegasun/}} while Echeclus is red, just as observed in our spectroscopic measurements. These colors, primarily due to their very large uncertainties, are also consistent with the spectral slopes determined from the X-Shooter spectra presented in Table \ref{tab:slopes}, with the colors of Echeclus and the coma respectively corresponding to slopes of approximately ${\sim}12\%$\phnm~and ${\sim}-10\%$\phnm. 

\subsection{Images: Coma Geometry}
During our observations of Echeclus, both its orbital velocity vector and the vector pointing from Echeclus to the Sun (the respective negatives of PsAMV and PsAng; see Fig. \ref{fig:f1} and Table \ref{tab:obsgeom}) were pointing roughly eastward on the sky. The angle between the Earth and Echeclus' orbital plane was also small, at ${\sim}0.3\degr$. From this vantage point we would expect that an extended coma would appear in the image to be oriented roughly westward if emitted with zero velocity relative to Echeclus and influenced only by forces of solar gravitation and the solar wind; this can easily be verified via the Finson--Probstein model \citep{1968ApJ...154..327F} implemented by \citet{2014acm..conf..565V}\footnote{\url{http://comet-toolbox.com/FP.html}}. The fact that the observed coma is oriented roughly southward indicates that it was emitted with some velocity, and is probably the remnant of a jet-like feature. The lack of broadening of this feature toward the anti-solar direction and the fact that it still remains six weeks post-outburst indicates that solar radiation pressure has not significantly affected its morphology, and likely indicates that the observed dust coma is comprised of large grains. 

\subsection{Images: Surface Brightness, \afr, and Coma Mass} \label{sec:phot}
\begin{figure*}
\includegraphics[scale=0.95]{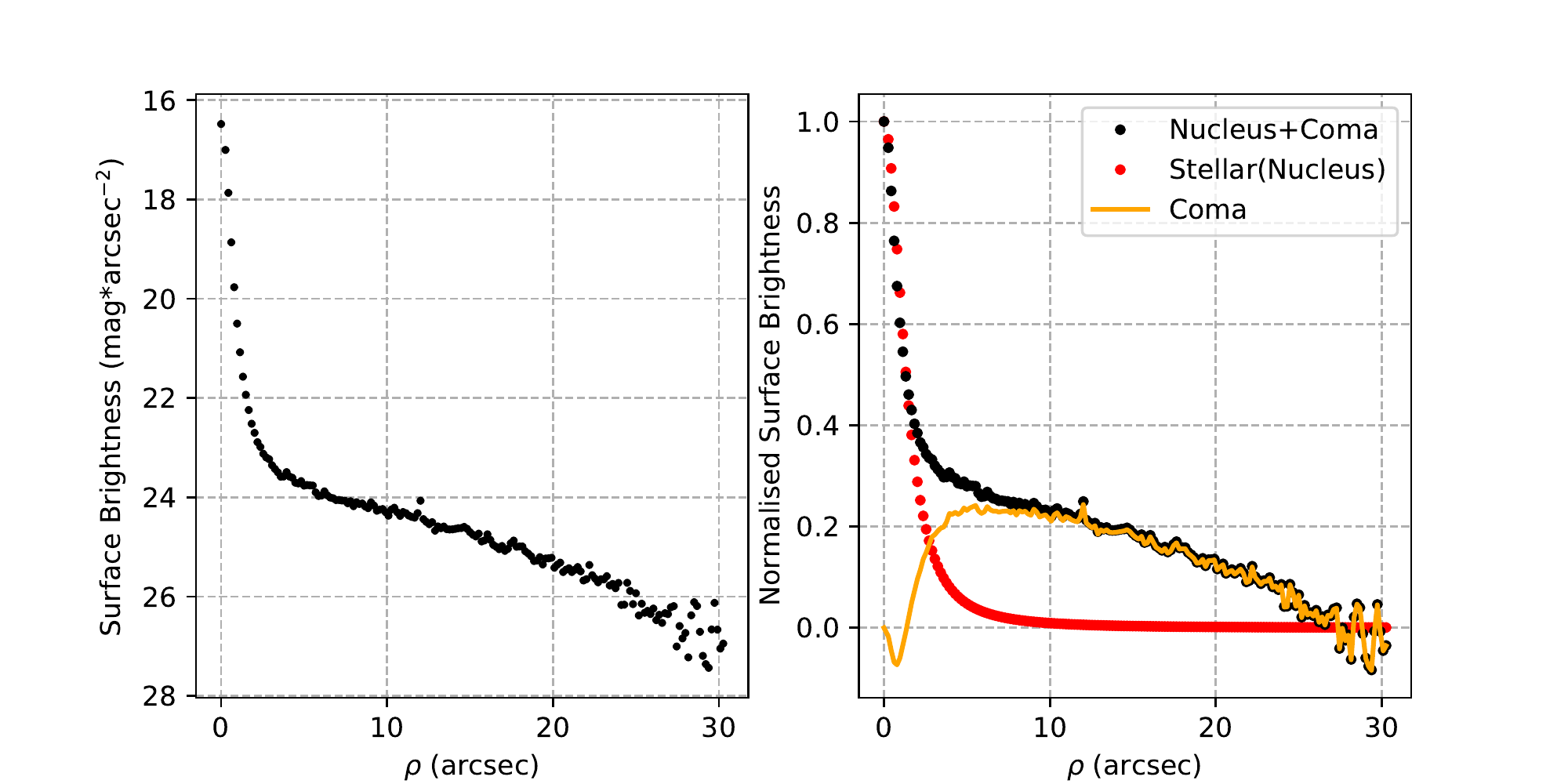}
\caption{Radially averaged surface brightness profiles (SBPs) of Echeclus produced from our single usable $r'$ image. Left: the SBP of echeclus given in magnitudes per arcsec$^{2}$. Error bars are present in this plot but are smaller than the plotted points. As explained in the text the photometric uncertainties should be considered larger than they are presented here, due to the poor flat-field calibration of the image. Right: the SBP of Echeclus normalized to one at its peak and to zero in the wings is shown with the normalized SPB calculated from a single fitted stellar PSF which we use as a proxy for the SBP of Echeclus' bare nucleus. The normalized SBP of the coma, produced by subtracting the stellar profile from the total profile, is also presented. The negative coma SBP at small $\rho$ is caused by the stellar SBP being wider than the total SBP close to the nucleus. As discussed in Section \ref{sec:phot}, the stellar SBP's greater broadness is an artifact of the non-sidereal tracking of Echeclus during the observation.\label{fig:f4}}
\end{figure*}

We determined the radially averaged surface brightness profile (SBP) of Echeclus by measuring the surface brightness of concentric rings centered on the photocenter of Echeclus; the rings had constant width and monotonically increasing radius, $\rho$. Here we only used the $r'$ image due to its higher precision zero-point calibration. The SBP of Echeclus plus its coma is displayed in Figure \ref{fig:f4}. To study the coma on its own we had to disentangle the SBP of the coma from that of the nucleus, which we have assumed to be the same as a stellar PSF. To remove the SBP of the nucleus we broadly followed the procedure described by \citet{2016Icar..271..314K}, first normalizing both the SBP of Echeclus and the SBP of the stellar PSF we previously saved. The SBPs were normalized to one at their peaks and to zero in their wings; in both cases the wings were normalized to the median surface brightness in the region $26\arcsec<\rho<30\arcsec$, after the points in that region were sigma clipped at 3$\sigma$. Beyond $\rho\sim26\arcsec$ the surface brightness of the coma blends into that of the background sky (see Figs. \ref{fig:f1} \& \ref{fig:f4}).

In each concentric ring of area, $A$ (measured in square arcseconds), the total flux, $F_{T}$, is the sum of the flux from the nucleus, $F_{N}$, and flux from the coma, $F_{C}$, and surface brightness is given by $\sigma=(F/A)$. Assuming that the flux contribution from the coma is negligible close to the nucleus, the ratio of nucleus surface brightness to that of total surface brightness in each ring is equal to the proportion of the total flux contributed by the nucleus, or $\sigma_{N}/\sigma_{T} = F_{N}/F_{T}$. Therefore in each ring, $F_{N} = F_{T}(\sigma_{N}/\sigma_{T})$ and $F_{C} = F_{T}(1 - (\sigma_{N}/\sigma_{T}))$. From here we can use the normalized Echeclus-plus-coma SBP and the normalized stellar SBP to determine the relative contributions to flux from the nucleus and coma in each ring (see Fig. \ref{fig:f4}). With a photometric aperture of radius 26\arcsec, the total brightness of Echeclus plus its coma was determined to be $r'_{T} = 15.98\pm0.05$ mag. Using the method described above, the flux of the nucleus and coma were disentangled and their brightnesses were respectively calculated to be $r'_{N} = 16.89 \pm 0.05$~mag and $r'_{C} = 16.62 \pm 0.05$~mag. 

Here we recognize that the coma's contribution to the flux is likely underestimated, and consequently the nucleus flux is overestimated. This is largely because our stellar SBP is broader than that of Echeclus at small values of $\rho$, and that flux from the near-nucleus coma is undersampled. Examination of the PSF of the star used to obtain our stellar SBP using SAOImageDS9 \citep{2003ASPC..295..489J} revealed that the star was trailed parallel to Echeclus' observed motion by around 0.19\arcsec~due to the non-sidereal tracking of Echeclus. The extent of this trailing is consistent with the discrepancy between the width of the stellar SBP and that of Echeclus at small $\rho$. It is also worth noting that with only a single stellar source from which to measure the stellar PSF, and the poor flat-fielding of the image, our estimated coma flux values and magnitudes very likely have larger uncertainties than those quoted here. As a result of these factors, the subsequent estimate for the mass of the dust ejected by the outburst is quoted merely at order-of-magnitude precision.

The parameter \afr~is designed to be a measure of the quantity of solar radiation reflected by cometary dust grains that is independent of observing geometry and the width of the measurement aperture, $\rho$ \citep{1984AJ.....89..579A}. \afr~has been calculated for previous outbursts of Echeclus, and has been used to estimate their dust production rates \citep{2008PASP..120..393B,2008A&26A...480..543R,2016MNRAS.462S.432R}. While \afr~can be a useful parameter, it has significant limitations, as explained in detail by \citet{2012Icar..221..721F}. The coma of Echeclus, as it appears in our observations, lacks steady-state outflow, spherical symmetry, and a $1/\rho$ column density dependence. This places our observations beyond the limitations within which \afr~should reasonably be applied, and renders any estimation of \afr~from our observations difficult to interpret. It is important to remember that \afr~should be used with care, and only within the context and limitations for which it was designed.

We estimated the solar flux reflected by a single dust grain in Echeclus' coma by rearranging the following equation \citep{1916ApJ....43..173R} to solve for $F_{D}$,
 
\begin{equation}
p = \frac{2.25\times10^{22}\Delta^{2}{r_{H}}^{2}}{{r_{D}}^{2}}\frac{F_{D}}{F_{\sun}},
\end{equation}
 
where $\Delta = 5.35$ au, $r_{H} = 6.35$ au, and $r'_{\sun} = -26.93$ \citep{2018ApJS..236...47W}, which corresponds to a flux, $F_{\sun} = 1.7\times10^{5}$~erg~s\textsuperscript{-1}~cm\textsuperscript{-2}. We assumed the grain's albedo, $p$, to be the same as that of Echeclus \citep[0.052;][]{2014A&A...564A..92D}, and that the grains observed in the image were in the Mie resonant scattering regime \citep[see][]{2012Hapkebook} with a radius of, $r_{D}=0.5~\micron$. From this we estimated the reflected flux from a single grain to be $F_{D} = 8.5\times10^{-35}$~erg~s\textsuperscript{-1}~cm\textsuperscript{-2}. The measured coma magnitude of $r'_{C} = 16.62$ corresponds to a flux, $F_{C} = 6.6\times10^{-13}$~erg~s\textsuperscript{-1}~cm\textsuperscript{-2}. We applied the phase angle correction of \citet{1998Icar..132..397S} to our flux estimate, although Echeclus was observed at a small phase angle of 0.9454\degr~and this correction is negligible within our large uncertainty margin. A lower limit to the number of dust grains in Echeclus' coma was calculated to be $N = F_{C}/F_{D}\sim8\times10^{21}$. Assuming spherical dust grains with a density of $\sim1\times10^{3}$~kg m\textsuperscript{-3} as measured by the GIADA instrument onboard \textit{Rosetta} \citep{2015Sci...347a3905R}, we estimate a lower limit on the total mass of Echeclus' dust coma to be of the order of $\sim10^{6}$~kg.
\newpage

\section{Discussion}
\subsection{Echeclus}
Despite evidence that cometary outbursts should change the reflectance properties of a centaur, including that gathered in-situ at comet 67P \citep{2016Icar..274..334F} and from laboratory experiments \citep{2016Icar..266..288P,2016Icar..267..154P}, we have not observed any change in the reflectance spectrum of Echeclus as a result of its 2016 outburst. We have not found any direct evidence to suggest the removal of an irradiated surface crust or the uncovering of fresh subsurface ices and silicates as predicted by \citet{2004A&26A...417.1145D} and \citet{2005Icar..174...90D}. This process should have manifested in the spectrum as the appearance of NIR absorption bands and the likely shallowing of the visual and NIR spectral gradients. Due to Echeclus' unchanging visual and NIR spectral gradients we also report no evidence for the blanketing of redder surface material by more neutral material falling back from the coma as predicted by \citet{2002AJ....123.1039J,2009AJ....137.4296J,2015AJ....150..201J}.

A simple explanation for the lack of change in Echeclus' spectrum may be that the outburst was not large enough to expose much of Echeclus's subsurface. Even on the much more active JFC 67P, it was found that the largest changes were localized to the most active regions of the nucleus surface \citep{2016Icar..274..334F}. If changes were localized to a small area they may not have been observable in our spatially unresolved measurements. It is unlikely that any large regions of exposed subsurface material could have been transformed by heat and irradiation from the Sun to be indistinguishable from the rest of Echeclus in only six weeks. Crystalline water ice is thermodynamically stable at $>$5 au and should have been observable if it was present on the Earth-facing side of Echeclus during our observations. Additionally, any irradiation mantle destroyed by activity could not be replaced by the irradiation of simple subsurface organics in the time between outburst and observations. \citet{2003crp4..791S} report that the production of a complex refractory organic crust requires ion irradiation doses of the order of 10$^{2}$~eV(16~amu)$^{-1}$ based on laboratory experiments. Under the solar wind at heliocentric distances of 5-10 au a surface takes ${\sim}10^{2}-10^{3}$ years to receive a total ion dose this high \citep{2012Icar..221...12K,2015Icar..248..222M}. Hence, if a large region of irradiated surface had been destroyed by the outburst to reveal an unirradiated subsurface, we would likely have detected it by changes in Echeclus' spectral slope. While the exposure of volatile ices such as carbon dioxide (CO$_{2}$) to seasonal heating on 67P has been shown to change the local surface composition on a timescale of weeks \citep{2016Sci...354.1563F}, our observations would imply a rapid disappearance of any potential deposit of volatile ices in any freshly exposed subsurface layers of Echeclus.

Another likely explanation for the lack of change in the nucleus' surface properties is that not enough comatic material fell to Echeclus' surface. If, as appears to be the case, the observed coma is a remnant of a cometary jet, and the dust ejected was traveling at a velocity much higher than Echeclus' escape velocity, it is likely that little of the coma material would have returned to Echeclus' surface under gravity. Assuming a spherical shape for the nucleus, a bulk density similar to that of 67P \citep[$\sim500$ kg m\textsuperscript{-3};][]{2015A&26A...583A..33P,2016Icar..277..257J}, and a diameter of around 60 km \citep{2013ApJ...773...22B,2014A&A...564A..92D}, Echeclus' escape velocity can be estimated at around 16~m~s\textsuperscript{-1}. Early images of the 2016 outburst were submitted to the joint comet image archive of the \textit{British Astronomical Association} and \textit{The Astronomer} by R. Miles and A. Watkins\footnote{\url{http://www.britastro.org/cometobs/174p/174p_20160905_rmiles.html}}; these images allowed them to estimate the expansion velocity of the coma of Echeclus to be $95\pm4$~m~s\textsuperscript{-1}. Hence, it appears that if this expansion velocity is representative of the velocity at which the majority of coma material was ejected, most of the dust ejected by the outburst would have been traveling too rapidly to be recaptured by Echeclus' low gravity, thus preventing the blanketing of the nucleus surface with blue comatic dust.

\subsection{Dust Coma}\label{sec:discoma}
The blue reflectance spectrum of Echeclus' dust coma, with a visual slope of {$-7.7\pm0.6\%$\phnm}, is unusual among active centaurs observed to date. Multiple active centaurs and JFCs, including Echeclus, have been reported to have comae that are more neutrally colored than the surfaces of their respective nuclei \citep{1989ApJ...347.1155A,1989AJ.....97.1766J,2003PASP..115..981B,2008PASP..120..393B,2008A&26A...480..543R,2009AJ....137.4296J,2017Icar..295...34F}; they are, however, typically still redder than solar at visual wavelengths. A notable exception, however, is (2060) Chiron. \citet{1991A&26A...241..635W} reported that Chiron's coma was both bluer than Chiron and bluer than the Sun when observed in 1990, with $B-V = 0.3\pm0.1$ \citep[solar $B-V = 0.653\pm0.003$;][]{2012ApJ...752....5R}. The coma was observed to be bluest closest to the nucleus where, due to Chiron's large size, the coma is gravitationally bound \citep{1995P&SS...43.1473F}. \citet{1991A&26A...241..635W} attributed the blue color of the coma to the presence of small non-geometrically scattering particles. In contradiction to this, however, \citet{1994A&26A...282..980F} reported that, based on models of Chiron's coma, the bound inner region of the coma was most likely dominated by larger grains of the order of 0.1~mm in size.

In typical cases of activity, non-geometric (or Rayleigh) scattering is often invoked as a cause for more neutral or blue colored comae, especially at larger cometocentric distances where smaller, non-geometrically scattering grains have been thrown off faster and further from the nucleus. \citet{2015AJ....150..201J}, however, argued that while optically small particles are numerically dominant in cometary comae, optically large particles dominate the scattering cross section, and the neutral/blue color of coma material is not dominated by small-particle scattering. In the case of Echeclus this interpretation is reinforced by the fact that our observations were performed six weeks post-outburst, after solar radiation pressure has likely dispersed the smallest grains. Those still observable are large enough to remain mostly unaffected by the influence of solar radiation pressure, and the lack of significant broadening of the observed coma toward the anti-solar direction (i.e. in the direction of PsAng in Fig. \ref{fig:f1}) supports this assessment. The apparent presence of large blue dust grains in the comae of both Echeclus and Chiron is intriguing; our observations and those of \citet{1991A&26A...241..635W} may be recording the effects of similar bluing processes happening in the comae of both these centaurs. Unfortunately, making direct comparisons between the dust present in the very different coma environments of Echeclus and Chiron is non-trivial, and in fact to do so may be entirely unreasonable.

If the color of Echeclus' dust coma is not a scattering effect it may be reflective of the dust's composition, with the best compositional candidate being carbon-rich organic matter. In the lab, amorphous carbon black, carbonaceous chondrites, and insoluble organic matter (IOM) from the Murchison meteorite have been observed to show dark, featureless, blue-sloped reflectance spectra \citep{1994Icar..107..276C,2011Icar..212..180C,2010JGRE..115.6005C}. Hydrocarbon and carbon-phase molecular fragments contained in dust particles released by active comets have also been measured or collected in-situ on multiple occasions: at 1P/Halley by the \textit{Vega} and \textit{Giotto} missions \citep{1986Natur.321..280K,1986Natur.321..336K}, at 81P/Wild by the \textit{Stardust} mission \citep{2006Sci...314.1720S}, and at 67P by multiple instruments onboard the \textit{Rosetta} spacecraft and its lander, \textit{Philae} \citep{2015Sci...349b0689G,2015Sci...349b0673W,2016Natur.538...72F}. Additionally, it was reported by \citet{2017MNRAS.469S.712B} that hydrocarbons are the dominant refractory material in 67P's dust grains, making up 50\% of the dust by mass. Hydrocarbons can become dehydrogenated when exposed to photonic and ionic radiation leading to a loss of their red reflectance properties in favor of a more neutral-blue spectrum \citep{1994Icar..107..276C,1998Icar..134..253M,2004Icar..170..214M}. This is borne out by studies of amorphous carbon collected at 81P which, if not primordial, was likely created by the breakup of more structured hydrocarbons by ion irradiation \citep{2008A&26A...485..743M,2009Icar..200..323B}. \citet{1994GeCoA..58.4503F} also found that in the coma of 1P, organic molecules were more abundant near to the nucleus compared to further away, suggesting that they were decomposing in the coma environment. Given that the dust observed and presented in this work had been exposed to direct solar irradiation for around six weeks, it is possible that complex hydrocarbons at the surface of the grains may have been dehydrogenated and broken up, such that the dust grains now have blue featureless reflectance properties dominated by relatively dehydrogenated amorphous carbon.

While composition is a tempting hypothesis to explain the dust's blue color, there are other ways to produce a blue slope that are worth noting. Laboratory experiments have observed that removal of the smallest particles from samples of CI (Ivuna-like) and CM (Mighei-like) carbonaceous chondrites\footnote{Meteorite classification is described in detail by \citet[][]{2006mess.book...19W}} produces a bluer reflectance spectrum compared to the original ensemble \citep{2010JGRE..115.6005C,2010LPI....41.1148H,2013LPI....44.1550C}. Observations of Murchison IOM have also shown strong bluing effects when observed at low phase angles \citep{2011Icar..212..180C}. Our observations were performed at a low phase angle of ${\sim}0.9\degr$ and are likely dominated by larger grains; hence, we cannot rule out these effects when considering the color of Echeclus' dust.

The way in which the dust's reflectance spectrum is blue in the visual, but levels out toward the NIR may have interesting implications for understanding the effects of blanketing that may take place as a result of cometary activity \citep{2002AJ....123.1039J}. Mixing enough of this blue dust into the red regolith of an unprocessed centaur would have a much stronger color neutralizing effect at visual wavelengths, while in the NIR the color may remain largely unchanged. This is something observed in the colours of centaurs, whereby their visual colours are bimodally distributed, but toward the NIR the color bifurcation is absent \citep[e.g.][]{2015A&26A...577A..35P}. Also, if the dust's color is indicative of a low albedo carbon-rich composition, deposition of this dust onto a centaur's surface would likely lower its albedo alongside neutralizing its color. Blanketing by carbon-rich dust of this kind could be a viable way to push a centaur from the higher albedo red color group into the lower albedo less-red color group \citep{2014ApJ...793L...2L}.

\section{Conclusions}
We have compared VNIR reflectance spectra of 174P/Echeclus covering the 0.4-2.1~\micron~range from 2014 while Echeclus was inert and six weeks after its 2016 outburst. The high S/N spectra were observed and reduced consistently across both epochs to ensure direct comparability between them. We did not observe any absorption features in Echeclus' spectrum following the outburst, nor have we observed any change in the visual or NIR spectral gradients at a statistically significant level. All of our measurements are broadly consistent with those published in previous works. The lack of change in the reflectance properties of Echeclus is likely due to the outburst not being strong enough to cause a change to the surface that would be observable from Earth. The apparent jet-like nature of the outburst suggests that most of the material ejected would easily escape Echeclus' low gravity, and very little comatic material would fall back to blanket the surface. If any significant change to Echeclus' surface has occurred, then it must be present at a location that was not Earth-facing during our observations.

A surprising result of this work is the observation of Echeclus' unusually blue dust coma, which from a reflectance spectrum we measured a visual slope of {$-7.7\pm0.6\%$\phnm} that levels out toward the NIR. From photometric measurements made post-outburst we also find that the \textit{g'-r'} color of the dust is also consistent with being bluer than solar, with $g'-r'=0.29\pm0.14$, corroborating our spectroscopic observations. It is unlikely that the blue color of the dust is caused by non-geometric scattering effects as the grains in the coma appear to be large; instead the color may be indicative of the dust's carbon-rich composition, but without discernible absorption features in the dust spectrum this cannot be confirmed. If the color of the dust is representative of its composition it is possible that deposition of enough of this material on the surface of a centaur may be able to neutralize the centaur's initially red optical color.

Based on our analysis of an \textit{r'} image obtained during our post-outburst observations, we have estimated a lower limit on the mass of the dust coma of $\sim10^{6}$~kg at the time of observation. 

\acknowledgments
We thank M.G.~Hyland for contributing to the data that went into our P97 DDT observing proposal, and O.~Hainaut for constructive referee comments that helped to improve the paper. We also thank the staff at Paranal Observatory for the great effort they put into performing our DDT observations; particular thanks go to G.~Beccari, J.~Pritchard, and B.~H\"au{\ss}ler. Spectra used in the P93 observing proposal were reduced using a pipeline written in IDL by G. Becker \citep[described by][]{2016A&26A...594A..91L}. T.S. acknowledges support from the Northern Ireland Department for the Economy, and the Astrophysics Research Centre at Queen's University Belfast. W.C.F. and A.F. acknowledge support from UK Science and Technology Facilities Council grant ST/P0003094/1. T.H.P. acknowledges support through FONDECYT Regular Grant No. 1161817 and CONICYT project Basal AFB-170002. This work is based on observations collected at the European Organisation for Astronomical Research in the Southern Hemisphere under ESO programs 093.C-0259(A) and 297.C-5064(A). This research made use of NASA's Astrophysics Data System Bibliographic Services, the JPL HORIZONS web interface (\url{https://ssd.jpl.nasa.gov/horizons.cgi}), and data and services provided by the IAU Minor Planet Center.

%

\vspace{5mm}
\facilities{ESO VLT(X-Shooter).}


\software{astropy \citep{2013A&A...558A..33A},  
          ESO Reflex \citep{2013A&A...559A..96F},
          matplotlib \citep{2007CSE...9..3H},
          numpy \citep{2011CSE...13..22V},
          SAOImageDS9 \citep{2003ASPC..295..489J},
          scipy \citep{2001SCIPY..J},
          SExtractor \citep{1996A&AS..117..393B}
          TRIPPy \citep{2016AJ....151..158F},
          The X-Shooter Pipeline \citep{2010SPIE.7737E..28M}
          }


\begin{thebibliography}{}

\bibitem[A'Hearn et al.(1989)]{1989ApJ...347.1155A} A'Hearn, M.~F., Campins, H., Schleicher, D.~G. \& Millis, R.~L.\ 1989, \apj, 347, 1155
\bibitem[A'Hearn et al.(1984)]{1984AJ.....89..579A} A'Hearn, M.~F., Schleicher, D.~G., Millis, R.~L., Feldman, P.~D. \& Thompson, D.~T.\ 1984, \aj, 89, 579
\bibitem[Abolfathi et al.(2018)]{2018ApJS..235...42A} Abolfathi, B., Aguado, D.~S., Aguilar, G., et al.\ 2018, \apjs, 235, 42
\bibitem[Alvarez-Candal et al.(2008)]{2008A&26A...487..741A} Alvarez-Candal, A., Fornasier, S., Barucci, M.~A., de Bergh, C. \& Merlin, F.\ 2008, \aap, 487, 741
\bibitem[Astropy Collaboration et al.(2013)]{2013A&A...558A..33A} Astropy Collaboration, Robitaille, T.~P., Tollerud, E.~J., et al.\ 2013, \aap, 558, A33
\bibitem[Bardyn et al.(2017)]{2017MNRAS.469S.712B} Bardyn, A., Baklouti, D., Cottin, H., et al.\ 2017, \mnras, 469, S712
\bibitem[Barucci et al.(2005)]{2005AJ....130.1291B} Barucci, M.~A., Belskaya, I.~N., Fulchignoni, M. \& Birlan, M.\ 2005, \aj, 130, 1291
\bibitem[Bauer et al.(2008)]{2008PASP..120..393B} Bauer, J.~M., Choi, Y.~J., Weissman, P.~R., et al.\ 2008, \pasp, 120, 393
\bibitem[Bauer et al.(2003a)]{2003PASP..115..981B} Bauer, J.~M., Fern\'andez, Y.~R. \& Meech, K.~J.\ 2003a, \pasp, 115, 981
\bibitem[Bauer et al.(2003b)]{2003Icar..166..195B} Bauer, J.~M., Meech, K.~M., Fern\'andez, Y.~R., et al.\ 2003b, \icarus, 166, 195
\bibitem[Bauer et al.(2013)]{2013ApJ...773...22B} Bauer, J.~M., Grav, T., Blauvelt, E., et al.\ 2013, \apj, 773, 22
\bibitem[Bertin \& Arnouts(1996)]{1996A&AS..117..393B} Bertin, E., \& Arnouts, S.\ 1996, \aaps, 117, 393
\bibitem[Boehnhardt et al.(2002)]{2002A&26A...395..297B} Boehnhardt, H., Delsanti, A., Barucci, A., et al.\ 2002, \aap, 395, 297
\bibitem[Brunetto et al.(2009)]{2009Icar..200..323B} Brunetto, R., Pino, T., Dartois, E., et al.\ 2009, \icarus, 200, 323
\bibitem[Choi et al.(2006a)]{IAUC8656} Choi, Y.~J., Weissman, P.~R. \& Polishook, D.\ 2006, \iaucirc, 8656
\bibitem[Choi et al.(2006b)]{CBET563} Choi, Y.~J., Weissman, P.~R., Chesley, S., et al.\ 2006, CBET, 563
\bibitem[Clark et al.(2010)]{2010JGRE..115.6005C} Clark, B.~E., Ziffer, J., Nesvorny, D., et al.\ 2010, JGR, 115, E06005
\bibitem[Cloutis et al.(1994)]{1994Icar..107..276C} Cloutis, E.~A., Gaffey, M.~J. \& Moslow, T.~F.\ 1994, \icarus, 107, 276
\bibitem[Cloutis et al.(2011)]{2011Icar..212..180C} Cloutis, E.~A., Hiroi, T., Gaffey, M.~J., Alexander, C.~M.~O'D. \& Mann, P.\ 2011, \icarus, 212, 180
\bibitem[Cloutis et al.(2013)]{2013LPI....44.1550C} Cloutis, E.~A., Hudon, P., Hiroi, T., et al.\ 2013, in Lunar and Planetary Science XLIV, Abstract \#1550
\bibitem[Delsanti et al.(2004)]{2004A&26A...417.1145D} Delsanti, A., Hainaut, O., Jourdeuil, E., et al.\ 2004, \aap, 417, 1145
\bibitem[Di Sisto \& Brunini(2007)]{2007Icar..190..224D} Di Sisto, R.~P. \& Brunini, A.\ 2007, \icarus, 190, 224
\bibitem[Dones et al.(1999)]{1999Icar..142..509D} Dones, L., Gladman, B., Melosh, H.~J. et al.\ 1999, \icarus, 142, 509
\bibitem[Doressoundiram et al.(2005)]{2005Icar..174...90D} Doressoundiram, A., Peixinho, N., Doucet, C., et al.\ 2005, \icarus, 174, 90
\bibitem[Duffard et al.(2014)]{2014A&A...564A..92D} Duffard, R., Pinilla-Alonso, N., Santos-Sanz, P., et al.\ 2014, \aap, 564, A92
\bibitem[Duncan et al.(2004)]{2004come.book..193D} Duncan, M., Levison, H.~F., \& Dones, L.\ 2004 in Comets II ed. Festou, M.~C. et al. (Tucson, AZ; Arizona University Press), 193
\bibitem[Fern\'andez et al.(2017)]{2017Icar..295...34F} Fern\'andez, J.~A., Licandro, J., Moreno, F., et al.\ 2017, \icarus, 295, 34
\bibitem[Fern\'andez(2009)]{2009P&26SS...57.1218F} Fern\'andez, Y.~R.\ 2009, \planss, 57, 1218
\bibitem[Filacchione et al.(2016a)]{2016Icar..274..334F} Filacchione, G., Capaccioni, F., Ciarniello, M., et al.\ 2016, \icarus, 274, 334
\bibitem[Filacchione et al.(2016b)]{2016Sci...354.1563F} Filacchione, G., Raponi, A., Capaccioni, F., et al.\ 2016, Sci, 354, 1563
\bibitem[Fink \& Rubin(2012)]{2012Icar..221..721F} Fink, U. \& Rubin, M.\ 2012, \icarus, 221, 721
\bibitem[Finson \& Probstein(1968)]{1968ApJ...154..327F} Finson, M.~L. \& Probstein, R.~F.\ 1968, \apj, 154, 327
\bibitem[Fitzsimmons et al.(2018)]{2018NatAs...2..133F} Fitzsimmons, A., Snodgrass, C., Rozitis, B.,  et al.\ 2018, NatAs, 2, 133
\bibitem[Flewelling et al.(2016)]{arXiv1612.05243} Flewelling, H.~A., Magnier, E.~A., Chambers, K.~C., et al.\ 2016, arXiv, 1612.05243, in preparation
\bibitem[Fomenkova et al.(1994)]{1994GeCoA..58.4503F} Fomenkova, M.~N., Chang, S. \& Mukhin, L.~M.\ 1994, GeCoA, 58, 4503
\bibitem[Fraser et al.(2016)]{2016AJ....151..158F} Fraser, W.~C., Alexandersen, M., Schwamb, M.~E., et al.\ 2016, \aj, 151, 158
\bibitem[Fray et al.(2016)]{2016Natur.538...72F} Fray, N., Bardyn, A., Cottin, H., et al.\ 2016, \nat, 538, 72
\bibitem[Freudling et al.(2013)]{2013A&A...559A..96F} Freudling, W., Romaniello, M., Bramich, D.~M. et al.\ 2013, \aap, 559, A96
\bibitem[Fulle(1994)]{1994A&26A...282..980F} Fulle, M.\ 1994, \aap, 282, 980
\bibitem[Fulle et al.(1995)]{1995P&SS...43.1473F} Fulle, M., Ortiz Gil, A. \& Pasian, F.\ 1995, Planet. Space Sci., 43, 1473
\bibitem[Gladman et al.(2008)]{2008ssbn.book...43G} Gladman, B., Marsden, B.~G. \& VanLaerhoven, C.\ 2008, in The Solar System Beyond Neptune, ed. Barucci, M.~A. et al. (Tucson, AZ; Arizona University Press), 43
\bibitem[Goesmann et al.(2015)]{2015Sci...349b0689G} Goesmann, F., Rosenbauer, H., Bredeh\"oft, J.~H., et al.\ 2015, Sci, 349, aab0689
\bibitem[Gomes et al.(2008)]{2008ssbn.book..259G} Gomes, R.~S., Fern\'andez, J.~A., Gallardo, T. \& Brunini, A.\ 2008 in The Solar System Beyond Neptune, ed. Barucci, M.~A. et al. (Tucson, AZ; Arizona University Press), 259
\bibitem[Guilbert et al.(2009)]{2009Icar..201..272G} Guilbert, A., Alvarez-Candal, A., Merlin, F., et al.\ 2009, \icarus, 201, 272
\bibitem[Hapke(2012)]{2012Hapkebook} Hapke, B.\ 2012, Theory of Reflectance and Emittance Spectroscopy (2nd ed; New York, NY: Cambridge University Press)
\bibitem[Hardorp(1980)]{1980A&A....91..221H} Hardorp, J.\ 1980, \aap, 91, 221
\bibitem[Hiroi et al.(2010)]{2010LPI....41.1148H} Hiroi, T., Jenniskens, P.~M., Bishop, J.~L. \& Shatir, T.\ 2010, in Lunar and Planetary Science XLI, Abstract \#1148
\bibitem[Horner et al.(2004)]{2004MNRAS.354..798H} Horner, J., Evans, N.~W. \& Bailey, M.~E.\ 2004, \mnras, 354, 798
\bibitem[Hunter(2007)]{2007CSE...9..3H} Hunter, J.~D.\ 2007, CSE, 9, 3
\bibitem[Jaeger et al.(2011)]{IAUC9213} Jaeger, M., Prosperi, E., Vollmann, W., et al.\ 2011, IAUC, 9213
\bibitem[Jewitt(2009)]{2009AJ....137.4296J} Jewitt, D.\ 2009 \aj, 137, 4296
\bibitem[Jewitt(2015)]{2015AJ....150..201J} Jewitt. D.\ 2015 \aj. 150, 201
\bibitem[Jewitt(2002)]{2002AJ....123.1039J} Jewitt, D.~C.\ 2002 \aj, 123, 1039
\bibitem[Jewitt \& Luu(1989)]{1989AJ.....97.1766J} Jewitt, D. \& Luu, J.\ 1989, \aj, 97, 1766
\bibitem[Jones et al.(2001)]{2001SCIPY..J} Jones, E., Oliphant, T., Peterson, P. et al.\ 2001, SciPy: Open source scientific tools for Python, http://www.scipy.org/
\bibitem[Jorda et al.(2016)]{2016Icar..277..257J} Jorda, L., Gaskell, R., Capanna, C., et al.\ 2016, \icarus, 277, 257
\bibitem[Joye \& Mandel(2003)]{2003ASPC..295..489J} Joye, W.~A. \& Mandel, E.\ 2003, in ASP Conf. Ser. 295, ADASS XII, ed. Payne, H.~E., Jedrzejewski, R.~I. \& Hook R.~N. (San Francisco, CA: ASP), 489
\bibitem[Ka\v nuchov\'a et al.(2012)]{2012Icar..221...12K} Ka\v nuchov\'a, Z., Brunetto, R., Melita, M. \& Strazzulla, G.\ 2012, \icarus, 221, 12
\bibitem[Kissel et al.(1986a)]{1986Natur.321..280K} Kissel, J., Sagdeev, R.~Z., Bertaux, J.~L., et al.\ 1986, \nat, 321, 280
\bibitem[Kissel et al.(1986b)]{1986Natur.321..336K} Kissel, J., Brownlee, D.~E., Buchler, K., et al.\ 1986, \nat, 321, 336
\bibitem[Kulyk et al.(2016)]{2016Icar..271..314K} Kulyk, I., Korsun, P., Rousselot, P., Afanasiev, V. \& Ivanova, O.\ 2016, \icarus, 271, 314
\bibitem[Lacerda et al.(2014)]{2014ApJ...793L...2L} Lacerda, P., Fornasier, S., Lellouch, E., et al.\ 2014, \apjl, 793, L2
\bibitem[Lamy \& Toth(2009)]{2009Icar..201..674L} Lamy, P. \& Toth, I.\ 2009, \icarus, 201, 674
\bibitem[Levison \& Duncan(1997)]{1997Icar..127...13L} Levison, H.~F. \& Duncan, M.~J.\ 1997, \icarus, 127, 13
\bibitem[L\'opez et al.(2016)]{2016A&26A...594A..91L} L\'opez, S., D'Odorico, V., Ellison, S.~L., et al.\ 2016, \aap, 594, A91
\bibitem[Luu \& Jewitt(1996)]{1996AJ....112.2310L} Luu, J.~X. \& Jewitt D.\ 1996, \aj, 122, 2310
\bibitem[Martayan et al.(2014)]{2014Msngr.156...21M} Martayan, C., Mehner, A., Beccari, G., et al.\ 2014, Msngr, 156, 21
\bibitem[Melita et al.(2015)]{2015Icar..248..222M} Melita, M., Ka\v nuchov\'a, Z., Brunetto, R. \& Strazzulla, G.\ 2015, \icarus, 248, 222
\bibitem[Miles et al.(2016)]{CBET4313} Miles, R., Camilleri, P., Birtwhistle, P. \& Gonzalez, J.~J.\ 2016, CBET, 4313
\bibitem[Modigliani et al.(2010)]{2010SPIE.7737E..28M} Modigliani, A., Goldoni, P., Royer, F., et al.\ 2010, Proc. SPIE, 7737, 28
\bibitem[Moffat(1969)]{1969A&A.....3..455M} Moffat, A.~F.~J.\ 1969, \aap, 3, 455
\bibitem[Moroz et al.(2004)]{2004Icar..170..214M} Moroz, L., Baratta, G., Strazzulla, G., et al.\ 2004, \icarus, 170, 214
\bibitem[Moroz et al.(1998)]{1998Icar..134..253M} Moroz, L.~V., Korochantsev, A.~V. \& W\"asch, R.\ 1998, \icarus, 134, 253
\bibitem[Mu\~noz Caro et al.(2008)]{2008A&26A...485..743M} Mu\~noz Caro, G.~M., Dartois, E. \& Nakamura-Messenger, K.\ 2008, \aap, 485, 743
\bibitem[Peixinho et al.(2015)]{2015A&26A...577A..35P} Peixinho, N., Delsanti, A. \& Doressoundiram, A.\ 2015, \aap, 577, 35
\bibitem[Peixinho et al.(2003)]{2003A&26A...410L..29P} Peixinho, N., Doressoundiram, A., Delsanti, A., et al.\ 2003, \aap, 410, L29
\bibitem[Perna et al.(2010)]{2010A&26A...510A..53P} Perna, D., Barucci, M.~A., Fornasier, S., et al.\ 2010, \aap, 510, 53
\bibitem[Poch et al.(2016a)]{2016Icar..266..288P} Poch, O., Pommerol, A., Jost, B., et al.\ 2016, \icarus, 266, 288
\bibitem[Poch et al.(2016b)]{2016Icar..267..154P} Poch, O., Pommerol, A., Jost, B., et al.\ 2016, \icarus, 267, 154
\bibitem[Preusker et al.(2015)]{2015A&26A...583A..33P} Preusker, F., Scholten, F., Matz, K.-D., et al.\ 2015, \aap, 583, A33
\bibitem[Ramirez et al.(2012)]{2012ApJ...752....5R} Ramirez, I., Michel, R., Sefako, R., et al.\ 2012, \apj, 752, 5
\bibitem[Rotundi et al.(2015)]{2015Sci...347a3905R} Rotundi, A., Sierks, H., Della Corte, V., et al.\ 2015, Sci, 347, aaa3905
\bibitem[Rousselot(2008)]{2008A&26A...480..543R} Rousselot, P.\ 2008, \aap, 480, 543
\bibitem[Rousselot et al.(2016)]{2016MNRAS.462S.432R} Rousselot, P., Korsun, P.~P., Kulyk, I., Guilbert-Lepoutre, A. \& Petit, J.-M.\ 2016, \mnras, 462, S432
\bibitem[Russell (1916)]{1916ApJ....43..173R} Russell, H.~N.\ 1916, \apj, 43, 173
\bibitem[Sandford et al.(2006)]{2006Sci...314.1720S} Sandford, S.~A., Al\'eon, J., Alexander, C.~M.~O'D., et al.\ 2006, Sci, 314, 1720
\bibitem[Schleicher et al.(1998)]{1998Icar..132..397S} Schleicher, D.~G., Millis, R.~L. \& Birch, P.~V.\ 1998, \icarus, 132, 397
\bibitem[Scotti et al.(2000)]{MPEC2000-E64} Scotti, J.~V., Gleason, A.~E., Montani, J.~L. \& Read, M.~T.\ 2000, MPEC 2000-E64
\bibitem[Seccull et al.(2018)]{2018ApJ...855L..26S} Seccull, T., Fraser, W.~C., Puzia, T.~H., Brown, M.~E. \& Sch\"onebeck, F.\ 2018, \apjl, 855, L26
\bibitem[Smette et al.(2015)]{2015A&26A...576A..77S} Smette, A., Sana, H., Noll, S., et al.\ 2015, \aap, 576, A77
\bibitem[Stansberry et al.(2008)]{2008ssbn.book..161S} Stansberry, J., Grundy, W., Brown, M., et al.\ 2008, in The Solar System Beyond Neptune, ed. Barucci, M.~A. et al. (Tucson, AZ; Arizona University Press), 161
\bibitem[Strazzulla et al.(2003)]{2003crp4..791S} Strazzulla, G., Cooper, J.~F., Christian, E.~R. \& Johnson, R.~E.\ 2003, CRPhy, 4, 791
\bibitem[Tegler et al.(2008)]{2008ssbn.book..105T} Tegler, S.~C., Bauer, J.~M., Romanishin, W. \& Peixinho, N.\ 2008, in The Solar System Beyond Neptune, ed. Barucci, M.~A. et al. (Tucson, AZ; Arizona University Press), 105
\bibitem[Tiscareno \& Malhotra(2003)]{2003AJ....126.3122T} Tiscareno, M.~S. \& Malhotra, R.\ 2003, \aj, 126, 3122
\bibitem[van der Walt et al.(2011)]{2011CSE...13..22V} van der Walt, S., Colbert, S.~C. \& Varoquaux, G.\ 2011, CSE, 13, 22
\bibitem[Vernet et al.(2011)]{2011A&A...536A.105V} Vernet, J., Dekker, H., D'Orico, S.  et al.\ 2011, \aap, 536, A105
\bibitem[Vernet et al.(2010)]{2010HiA....15..535V} Vernet, J., Kerber, F., Mainieri, V., et al.\ 2010, HiA, 15, 535
\bibitem[Vincent(2014)]{2014acm..conf..565V} Vincent, J.\ 2014, in Asteroids, Comets, Meteors 2014, ed. K. Muinonen et al. (Helsinki, Finland; University of Helsinki), 565
\bibitem[Weisberg et al.(2006)]{2006mess.book...19W} Weisberg, M.~K., McCoy, T.~J. \& Krot, A.~N.\ 2006, in Meteorites and the Early Solar System II, ed. Lauretta, D.~S. \& McSween, H.~Y.~Jr (Tucson, AZ; Arizona Univ. Press), 19
\bibitem[Weissman et al.(2006)]{BAAS38551} Weissman, P.~R., Chesley, S.~R., Choi, Y.~J. et al.\ 2006, \baas, 38, 551
\bibitem[West(1991)]{1991A&26A...241..635W} West, R.~M.\ 1991, \aap, 241, 635
\bibitem[Wierzchos et al.(2017)]{2017AJ....153..230W} Wierzchos, K., Womack, M. \& Sarid, G.\ 2017, \aj, 153, 230
\bibitem[Willmer(2018)]{2018ApJS..236...47W} Willmer, C.~N.~A.\ 2018, \apjs, 236, 47
\bibitem[Wright et al.(2015)]{2015Sci...349b0673W} Wright, I.~P., Sheridan, S., Barber, S.~J., et al.\ 2015, Sci, 349, aab0673

\end{thebibliography}
\end{document}